\documentclass[sigconf]{acmart}
\usepackage{pbalance}
\usepackage{enumitem}
\usepackage{booktabs}
\usepackage{graphicx}
\usepackage{array}
\usepackage{lineno}
\usepackage{multirow}
\usepackage{hhline}
\usepackage{makecell}
\usepackage{subfigure}
\usepackage[marginal]{footmisc}

\AtBeginDocument{%
  \providecommand\BibTeX{{%
    \normalfont B\kern-0.5em{\scshape i\kern-0.25em b}\kern-0.8em\TeX}}}

\copyrightyear{2025}
\acmYear{2025}
\setcopyright{acmlicensed}
\acmConference[CIKM '25]{Proceedings of the 34th ACM International Conference on Information and Knowledge Management}{November 10--14, 2025}{Seoul, Republic of Korea}
\acmBooktitle{Proceedings of the 34th ACM International Conference on Information and Knowledge Management (CIKM '25), November 10--14, 2025, Seoul, Republic of Korea}
\acmDOI{10.1145/3746252.3761539}
\acmISBN{979-8-4007-2040-6/2025/11}

\begin{document}

\title{
You Only Evaluate Once: A Tree-based Rerank Method at Meituan
}


\author{Shuli Wang}
\authornote{Corresponding author.}
\affiliation{%
  \institution{Meituan}
   \city{Chengdu}
  \country{China}
}
\email{wangshuli03@meituan.com}

\author{Yinqiu Huang}
\affiliation{%
\institution{Meituan}
   \city{Chengdu}
  \country{China}
  }
\email{huangyinqiu@meituan.com}

\author{Changhao Li}
\affiliation{%
  \institution{Meituan}
   \city{Chengdu}
  \country{China}
}
\email{lichanghao@meituan.com}

\author{Yuan Zhou}
\affiliation{%
\institution{Meituan}
   \city{Chengdu}
  \country{China}
  }
\email{zhouyuan22@meituan.com}

\author{Yonggang Liu}
\affiliation{%
\institution{Meituan}
   \city{Chengdu}
  \country{China}
  }
\email{liuyonggang02@meituan.com}

\author{Yongqiang Zhang}
\affiliation{%
\institution{Meituan}
   \city{Chengdu}
  \country{China}
  }
\email{zhangyongqiang08@meituan.com}

\author{Yinhua Zhu}
\affiliation{%
\institution{Meituan}
   \city{Chengdu}
  \country{China}
  }
\email{zhuyinhua@meituan.com}

\author{Haitao Wang}
\affiliation{%
\institution{Meituan}
   \city{Chengdu}
  \country{China}
  }
\email{wanghaitao13@meituan.com}

\author{Xingxing Wang}
\affiliation{%
\institution{Meituan}
   \city{Beijing}
  \country{China}
  }
\email{wangxingxing04@meituan.com}

\renewcommand{\shortauthors}{Shuli Wang et al.}

\begin{abstract}


Reranking plays a crucial role in modern recommender systems by capturing the mutual influences within the list. Due to the inherent challenges of combinatorial search spaces, most methods adopt a two-stage search paradigm: a simple General Search Unit (GSU) efficiently reduces the candidate space, and an Exact Search Unit (ESU) effectively selects the optimal sequence. 
These methods essentially involve making trade-offs between effectiveness and efficiency, while suffering from a severe \textbf{inconsistency problem}, that is, the GSU often misses high-value lists from ESU. To address this problem, we propose YOLOR, a one-stage reranking method that removes the GSU while retaining only the ESU.
Specifically, YOLOR includes: (1) a Tree-based Context Extraction Module (TCEM) that hierarchically aggregates multi-scale contextual features to achieve "list-level effectiveness", and (2) a Context Cache Module (CCM) that enables efficient feature reuse across candidate permutations to achieve "permutation-level efficiency".
Extensive experiments across public and industry datasets validate YOLOR's performance and we have successfully deployed YOLOR on the Meituan food delivery platform.

\end{abstract}





\maketitle

\section{Introduction}
\begin{figure}[h]
\centering
\includegraphics[width=0.7\linewidth, height=1\textheight, keepaspectratio]{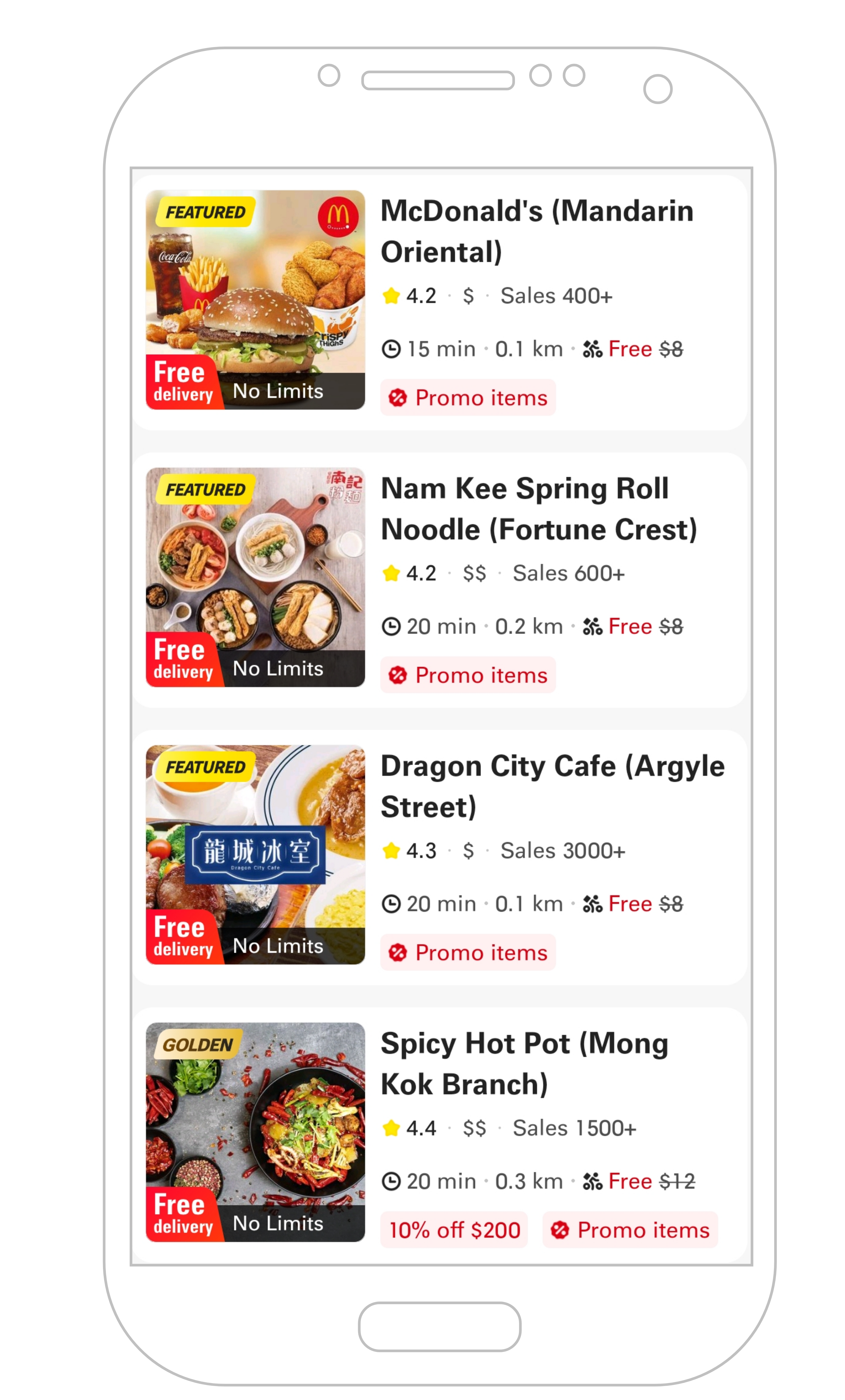}
\caption{List recommendations on Meituan food delivery platform.
}
\label{fig: demo}
\end{figure}

E-commerce platforms, such as Meituan and Taobao, need to provide users with personalized services from millions of items. As shown in Figure \ref{fig: demo}, list recommendation is the main display form on Meituan food delivery platform.
To improve recommendation efficiency, personalized recommendation systems generally include three stages: matching,  ranking, and reranking. The ranking models evaluate the recommended items respectively, focusing on feature interactions \citeN{W&D, deepfm, xdeepfm},
user preference modeling \citeN{din, dien, sim}, and so on. However, ranking methods ignore the crucial mutual influence among contextual items. 
Research \citeN{burges2010ranknet, listnet, ai2018learning, pang2020setrank} indicates that optimizing a listwise utility during the reranking stage is a more advantageous strategy, as it capitalizes on the mutual influences between items within the list to enhance overall performance.

The key challenge in reranking is exploring optimal lists within the vast permutation space. Existing reranking methods can be classified into two categories \citeN{grn, pier}. The first category is generator-based methods. These methods generate the list by some heuristic strategy \citeN{seq2slate, gong2022real, zhuang2018globally}, or generate a suboptimal list with greedy local sight \citeN{grn}. 
Despite considering the context, these methods fail to ensure the quality of the generated lists and obtain optimal results, due to issues such as evaluation-before-reranking problems \cite{xi2021context}.
The second category is evaluator-based methods \citeN{feng2021revisit, xi2021context, pier}. These methods try to evaluate every possible permutation to get the optimal list.
However, due to the strict time constraints in online systems, most existing evaluator-based methods use a two-stage architecture, which adds a filtering submodel \citeN{pier, feng2021revisit} or generating submodel \citeN{NAR4Rec} as the General Search Unit (GSU) before the Exact Search Unit (ESU) to reduce the size of the candidate space.

However, existing two-stage reranking methods fall into trade-off and face two significant issues. \textbf{Firstly, the inherent inconsistency problem between the two stages limits the effectiveness of the GSU.} GSU is expected to search high-value lists to ESU. However, due to strict time constraints, the GSU can only be designed simply but not precisely. Although some recent generative reranking methods \citeN{grn,NAR4Rec,dcdr} leverage the ESU as a teacher to guide the GSU, GSU's inconsistency issues remain serious \citeN{pier,nlgr} and lead to unsatisfactory hit ratio due to the vast permutation space. The inconsistency problem arises from the trade-offs inherent in the two-stage model \cite{chang2023twin} which is challenging to solve.
\textbf{Secondly, lacking of multi-scale contextual information limits the effectiveness of ESU.} As shown in Figure \ref{fig: demo}, in the list recommendation tasks, the context information between items is complex, and the impact windows between items may span 2, 4 or even the entire list.
\begin{figure}[h]
\centering
\includegraphics[width=0.95\linewidth, height=1\textheight, keepaspectratio]{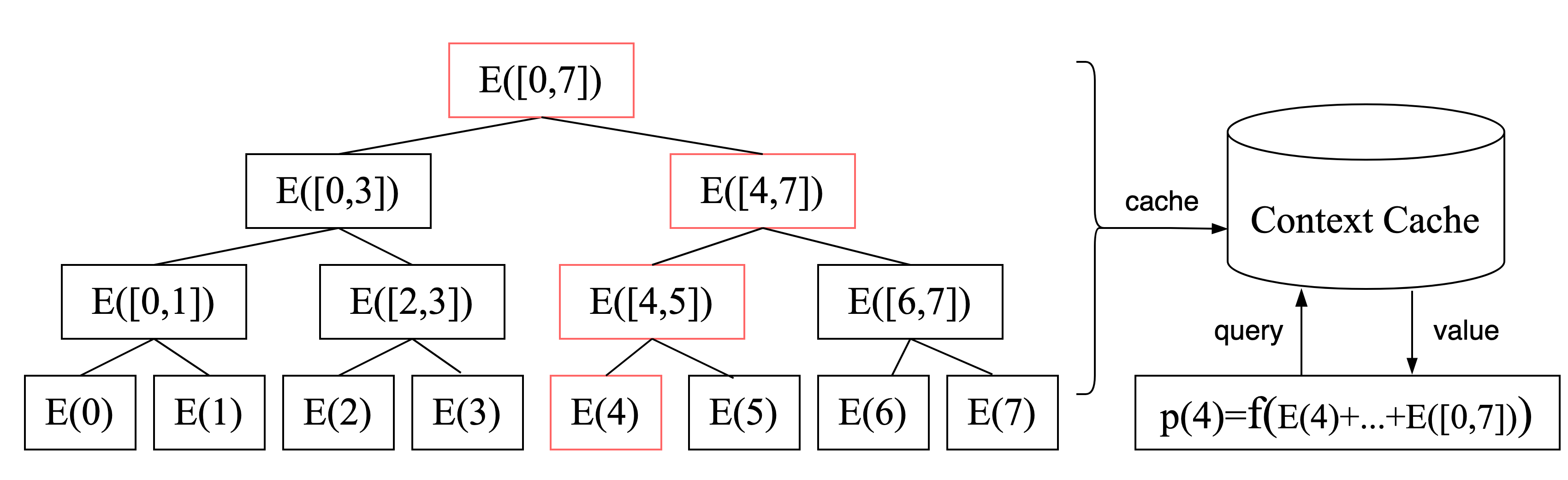}
\caption{Demo of YOLOR. When evaluating the 4th item, we can obtain its precise score by combining its multi-scale contextual information and achieve efficient reuse through caching. 
}
\label{fig: 2}
\end{figure}

To resolve the aforementioned issues, we propose a tree-based \underline{R}erank framework that \underline{Y}ou \underline{O}nly eva\underline{L}uate \underline{O}nce, termed YOLOR, which removes the GSU while retaining only the ESU.
We resolve the efficiency-effectiveness trade-off by simultaneously addressing "permutation-level efficiency" and "list-level effectiveness".
As shown in Figure \ref{fig: 2}, for effectiveness at the list level, we designed a tree-based context extraction module to capture contextual information at different scales. For efficiency at the permutation level, we developed a context caching module to enable efficient reuse of multi-scale contextual information across lists. It is an extremely time-saving method that allows you to predict all permutations. 

The main contributions of our work are summarized as follows:
\begin{itemize}

\item We propose a novel framework that resolves the efficiency-effectiveness trade-off through simultaneous optimization of "permutation-level efficiency" and "list-level effectiveness". To the best of our knowledge, we are the first to introduce this concept.

\item Based on this idea, we design the YOLOR model, which contains the TCEM module to extract multi-scale contextual information and the CCM module to reduce repeated calculations.

\item We conduct extensive experiments on both offline and real-world industrial datasets from Meituan. Experimental results demonstrate the effectiveness of YOLOR. It is notable that YOLOR has been deployed in Meituan food delivery platform and has achieved significant improvement under various metrics.

\end{itemize}

\section{Related Work}

In recommendation systems, the core of the reranking stage lies in modeling the context and selecting the optimal list from the permutation space. Existing research on reranking can be systematically classified into two principal categories\cite{gfn}: generator-based methods \citeN{prm, mir, grn} and evaluator-based methods \citeN{feng2021revisit, pier}.

Generator-based methods directly generate one list as output by capturing the mutual influence among items. Therefore, these generator-based methods are also called one-stage methods \citeN{pier, NAR4Rec}. Initially, these methods directly used behavioral logs as training guides. For instance, Seq2slate \cite{seq2slate} utilizes pointer-network and MIRNN \cite{zhuang2018globally} utilizes GRU to determine the item order one-by-one. Methods such as PRM \cite{prm} and DLCM \cite{ai2018learning} take the initial ranking list as input, use RNN or self-attention to model the context-wise signal, and output the predicted value of each item. Such methods bring an evaluation-before-reranking problem \cite{xi2021context} and lead to sub-optimum. Similarly, methods such as EXTR \cite{extr} estimate the predicted Click-Through Rate (pCTR) of each candidate item on each candidate position, which are substantially point-wise models and thus limited in extracting exact context. MIR \cite{mir} capturing the set2list interactions by a permutation-equivariant module.
Since it is difficult to achieve permutation space optimization in supervised training, some generator-based methods using evaluators \citeN{listcvae, grn} have become popular in recent years. For instance, ListCVAE \cite{listcvae} utilizes conditional variational autoencoders (CVAE) to capture the positional biases of items and the interdependencies within the list distribution. GRN \cite{grn} proposes an evaluator-generator framework to replace the greedy strategy, but it can't avoid the evaluation-before-reranking problem \cite{xi2021context} because it takes the rank list as input to the generator. DCDR \cite{dcdr} introduces diffusion models into the reranking stage and presents a discrete conditional diffusion reranking framework. NAR4Rec \citeN{NAR4Rec} uses a non-autoregressive generative model to speed up sequence generation. However, this paradigm heavily depends on the accuracy of the evaluator which makes it less promising in industrial recommendation tasks.

Evaluator-based methods try to evaluate every possible permutation through a well-designed context-wise model. Due to the strict time constraints in online systems, most existing evaluator-based methods use a two-stage architecture, which adds a filtering stage \citeN{pier, feng2021revisit} or generating stage \citeN{NAR4Rec} before the evaluating stage to reduce the size of the candidate set. For instance, PRS \cite{feng2021revisit} adopts beam-search to generate a few candidate permutations first, and score each permutation through a permutation-wise ranking model. PIER \cite{pier} applies SimHash \citeN{charikar2002similarity, chen2021end,manku2007detecting} to select top-K candidates from the full permutation. In industrial recommendation systems, the concept of evaluator-based methods is broader, as any generator-based method can be utilized in the first stage and form a multi-channel retrieval framework \citeN{NAR4Rec}. However, the inconsistency problem between the two stages limits the effectiveness of the model.
    
\begin{figure*}[h]
\centering
\includegraphics[width=\textwidth]{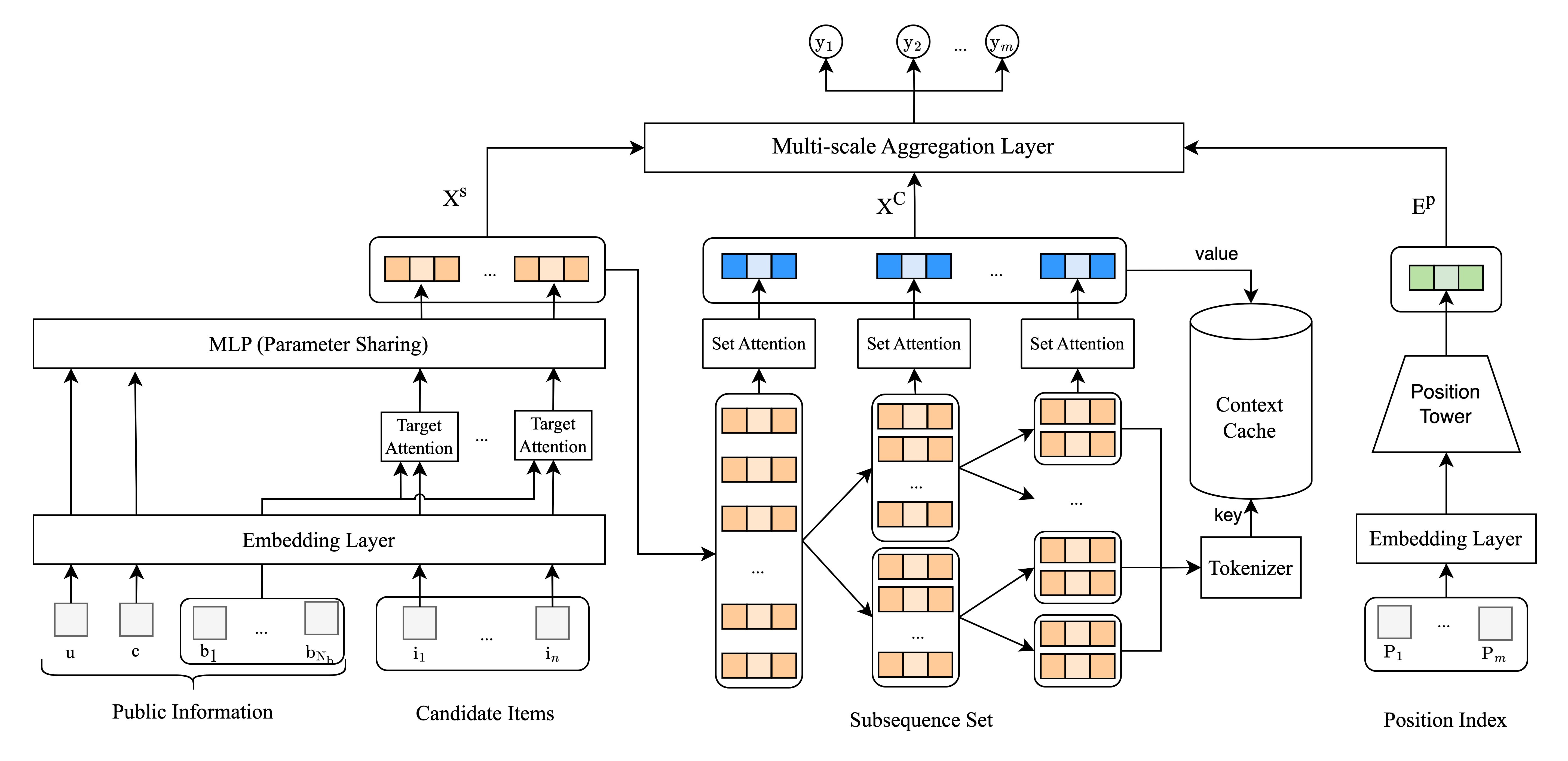}
\caption{The overall architecture of YOLOR.}
\label{fig:yolor}
\end{figure*}

\section{Problem Definition}

Let $ U = \{ u_1, u_2, \ldots, u_{|U|} \} $  represent a set of $|U|$ users which consist of some profile features (e.g. user ID, gender, age) and recent interaction history. For each user $u$, given a candidate set with $n$ items $X = \{ x_1, x_2, \ldots, x_n \} $, the goal of reranking is to propose an ordered list with $m$ ($m \leq n$) items $L = \{ x_1, x_2, \ldots, x_m \} $ from $\mathcal{O}(A_n^m)$ candidate space $\mathcal{L}$ and maximizes the listwise reward $\mathcal{R}(u, L)$: 
\begin{equation}\label{eq:1}
L^*=\mathop{\arg\max}\limits_{L}\mathcal{R}(u, \mathcal{L}),
\end{equation}
where $\mathcal{R}(u, L) = \sum^m_{i=1} \mathcal{R}(u, \mathbf{x}_i)$.

\section{Proposed Method}

In this section, we will introduce the structure of YOLOR in detail. As shown in Figure \ref{fig:yolor}, YOLOR mainly includes three modules, namely the Item-level Representation Module (IRM), the Tree-based Context Extraction Module (TCEM) for modeling contextual information, and Context Cache Module (CCM). We will introduce them in detail in the following subsections.

\subsection{Item-level Representation Module}\label{section:IRM}

We use the Item-level Representation Module (IRM) to generate the semantic embedding of each candidate item from the raw input. First, we use an embedding layer to get the embedding of the original input. We denote the embeddings of the original input user profile features, context features, user behaviors sequence and candidate items features as $\mathbf{e}^u$, $\mathbf{e}^c$, $\mathbf{E}^b \in \mathbb{R}^{N_b \times D}$ and $\mathbf{X} \in \mathbb{R}^{n \times D}$ respectively, where $N_b$ and $n$ are the number of user behaviors and candidate items, $D$ is the dimension of the embedding layer. Then, we use a target attention unit to encode the interaction between the historical behaviors of the user and the corresponding item:
\begin{equation}
\mathbf{x}'_i=\mathrm{Attention}(\mathbf{X}_i, \{\mathbf{E}^b_j\}_{j=1}^{N_b}), \forall i \in [n],
\end{equation}
where $\mathbf{X}_i \in \mathbb{R}^{D}$ is the $i$-th candidate item in $\mathbf{X}$.

Then, we use MLP as a simple feature crosses unit to extract the semantic embedding of each candidate item:
\begin{equation}\label{eq:4}
\mathbf{x}^s_i=\mathrm{MLP}\left(\mathbf{x}'_i||\mathbf{e}^u||\mathbf{e}^c\right), \forall i \in [n],
\end{equation}
where || represents concatenate operate. 

For ease of notation, we can also write the semantic embeddings for items in matrix form, each row of which represents one item in the sequence, i.e.,
\begin{equation}\label{formula:x_s}
\mathbf{X}^s=[\mathbf{x}^s_1;\mathbf{x}^s_2;...;\mathbf{x}^s_n]^\top.
\end{equation}
Note that this is just the simplest implementation. IRM is only $\mathcal{O}(n)$ complexity rather than $\mathcal{O}(A_n^m)$, so it can be more complex.

Unlike other reranking methods that solely rely on the predicted scores from ranking models, we directly deploy the ranking model as IRM in our online system, which enhances the consistency of the recommendation pipeline. The details will be discussed in Section \ref{subsection:5.6}.

\subsection{Tree-based Context Extraction Module}\label{section:TCEM}

We design the Tree-based Context Extraction Module (TCEM) to model multi-scale contextual relationships. For the $t$-th item $x_t$ in the candidate list $L$, we construct multi-scale subsequences to extract $x_t$'s local and global contextual information. First, We split the list $L$ constantly until there are only two items left in the subsequence, then we obtain the set of all subsequence denoted as:
\begin{equation} \label{equation:6}
C = \{L, L_{1,m/2}, L_{m/2+1,m}, L_{1,m/4}, \ldots, L_{t,t+1}, \ldots, L_{m-1,m}\},
\end{equation}
where $L_{l,r}$ represents the subsequence of list $L$ from index $l$ to index $r$, denoted as $L_{l,r}=[\mathbf{x}^s_l;\mathbf{x}^s_{l+1};...;\mathbf{x}^s_r]$. 

Then, we collect all subsequences containing $x_t$, denoted as $C_t = \{L, \ldots, L_{t,t+1}\}$, and perform contextual information extraction for each subsequence in $C_t$:
And then we input each subsequence in $C_t$ into a Self-Attention layer (SA) \cite{vaswani2017attention} to extract contextual information $\mathbf{e}_{l,r} \in \mathbb{R}^{D}$:

\begin{equation} \label{eq:7}
\begin{aligned}
&\mathbf{e}_{(1)} = \mathbf{e}_{1,m} = \mathrm{SA}(\mathbf{x}^s_1 || \mathbf{x}^s_{2} || ... || \mathbf{x}^s_m),
\\
&\mathbf{e}_{(2)} = \mathbf{e}_{l,r} = \mathrm{SA}(\mathbf{x}^s_l || \mathbf{x}^s_{l+1} || ... || \mathbf{x}^s_r),
\\
&...
\\
&\mathbf{e}_{(\log_{2}{m})} = \mathbf{e}_{t,t+1} = \mathrm{SA}(\mathbf{x}^s_t || \mathbf{x}^s_{t+1}),
\end{aligned}
\end{equation}
where || represents concatenate operate. 
Note that the SA layer in the above formula does not have position encoding, which can increase reusability and reduce calculations, also termed Set Attention. Through Eq. \ref{eq:7}, we can obtain multi-scale contextual relationships associated with $x_t$, denoted as $\mathbf{X}^C_t=[\mathbf{e}_{(1)};\mathbf{e}_{(2)};\ldots;\mathbf{e}_{(\log_{2}{m})}] \in \mathbb{R}^{D \cdot\log_{2}{m}}$.

Finally, we use a parameter-sharing fully connected layer to predict the list-wise pCTR of each item in each permutation. Taking the $t$-th item in permutation as an example still, the inputs consist of three parts: absolute position representation of $t$-th position $\mathbf{E}^p_t \in \mathbb{R}^{D}$, item representation of $t$-th position $\mathbf{X}^s_t \in \mathbb{R}^{D}$, and context representation of $t$-th position $\mathbf{X}^C_t$. Then the list-wise pCTR of the $t$-th item is predicted as follows:

\begin{equation}\label{eq:8}
\hat{y}_t = \sigma \left( \mathrm{FC}(\mathbf{E}^p_t || \mathbf{X}^s_t || \mathbf{X}^C_t) \right),
\end{equation}
where $\sigma$ is the Sigmoid Function. The score of each list is easily obtained by summing the output list-wise pCTR:
\begin{equation}\label{eq:9}
\hat{y}_L = \sum{(\hat{y}_1,\hat{y}_2,...,\hat{y}_t,...,\hat{y}_m)}.
\end{equation}


\subsection{Context Cache Module}

We have obtained the item-level representation and contextual representation of the $t$-th item $x_t$ through the aforementioned IRM and TCEM modules, respectively. However, due to strict time constraints, such a complex method cannot be applied directly to the permutation space. Therefore, we further design a Context Cache Module (CCM) to enable efficient reuse, which only requires simple matrix operation.

First, we generate the set of all subsequences (including candidate items $\mathbf{X}$) in the permutation space. To avoid notational confusion, we still denote this set as $C$. Clearly, the size of $C$ is $|C|=C_n^m + C_n^{m/2} + ... + C_n^{1}$. Then, based on Eq. \ref{eq:7}, we extract contextual information to obtain the multi-scale context representation matrix $\mathbf{X}^C \in \mathcal{R}^{|C| \times D}$. 

Next, to evaluate all candidate lists in the permutation space $\mathcal{L} \in \mathcal{R}^{A_n^m \times m}$, we gather the required contextual information for $\mathcal{L}$ from $\mathbf{X}^C$, denoted as:

\begin{equation}\label{eq:gather}
\mathbf{X}^C_\mathcal{L} = \mathrm{tf.gather}(\mathbf{X}^C, M_{indices}),
\end{equation}
where tf.gather \footnote{https://www.tensorflow.org/api\_docs/python/tf/gather} is Tensorflow's gather operator. And $M_{indices} \in \mathcal{R}^{A_n^m \times m \times \log_{2}{m}}$ is a request-independent matrix as long as m and n are fixed.

Then, based on Eq. \ref{eq:8} and Eq. \ref{eq:9}, we calculate the list-wise score for each candidate list $\hat{y}_{\mathcal{L}} \in \mathcal{R}^{A_n^m}$:
\begin{equation}\label{eq:final}
\hat{y}_{\mathcal{L}} = \mathrm{reduce\_sum}\left(\sigma\left(\mathrm{FC}(\mathrm{tile}(E^p)||\mathbf{X}^C_\mathcal{L})\right), \mathrm{axis=-1}\right).
\end{equation}

Finally, based on Eq. \ref{eq:1}, we select the optimal list:
\begin{equation}\label{eq:11}
L^*=\mathop{\arg\max}\limits_{L}\hat{y}_{\mathcal{L}}.
\end{equation}
It should be noted that the score of each list can be conveniently adjusted according to business needs, such as CVR (Conversion Rate) and GMV (Gross Merchandise Volume).

\subsection{Model Complexity} 

We perform a model complexity analysis of YOLOR to illustrate that our model meets the standards for online deployment. As mentioned, reranking models face serious challenges of $\mathcal{O}(A_n^m)$ candidate space. YOLOR uses the CCM module to reduce calculations. The computational complexity of IRM is $\mathcal{O}(n)$. 
The original computational complexity of TCEM is $\mathcal{O}(A_n^m * \log_{2}{m})$. 
But through the CCM module, we can fully reuse the context information. 
The additional space complexity required by CCM is $\mathcal{O}(C_n^m, C_n^{m/2}, ..., C_n^{2}, C_n^{1})$.
Take $n=m=8$ as an example, there are a total of $\mathcal{O}(A_8^8)=40320$ lists in the permutation space, and CCM requires storing only $\mathcal{O}(C_8^8, C_8^{4}, C_8^{2}, C_8^{1})=107$ context embeddings, a few amount of storage is sufficient to enable rapid computations at the permutation level.
The remaining only $\mathcal{O}(A_n^m)$ complexity in YOLOR is Eq. \ref{eq:final}, and the computational cost of Eq. \ref{eq:gather} is negligible compared to that of Eq. \ref{eq:final}. We design the prediction layer as a single FC layer, resulting in minimal computational complexity. Therefore, the complexity of YOLOR is acceptable for online serving.

\subsection{Model Training} 

We train YOLOR using real data collected from online logs. The input is the features of the recommended advertisement lists exposed in reality online, and the advertising return situation, including exposure, click, conversion, and other performance indicators, is used as the label to supervise the training of YOLOR. 
First, we use cross-entropy loss to train YOLOR, the loss of each list is calculated as follows:

\begin{equation}
\mathcal{L}_{ce}  = -\frac{1}{m} \sum_{t=1}^m \left( y_ {t}  \log  ( \widehat {y}_{t}) + (1-  y_ {t}  )  \log  {(1-\widehat {y}_{t}} ) \right),
\end{equation}
where subscript $t$ is the index of displayed items, $y_t$ represents the real label, $\widehat{y}_{t}$ represents the predicted value, $m$ is the length of exposed list.
Additionally, we set a dropout rate to mask contextual information randomly.

Then, to enhance the context extraction ability, we propose GBPR loss (Group BPR \cite{rendle2012bpr} loss) to enhance the contrast of the positive and negative samples within each exposure list:

\begin{equation}\label{eq:gbpr}
\mathcal{L}_{gbpr}  = -\frac{1}{m} \sum_{t=1}^m \log \left(    \mathrm{sgn}(y_i - y_j)( \widehat {y}_{i} -\widehat {y}_{j} ) \right),
\end{equation}
where subscript $i$ and $j$ are the indexes of displayed items. And $\mathrm{sgn}(x)$ represents sign function, $\mathrm{sgn}(x)=0$ if $x=0$, $\mathrm{sgn}(x)=1$ if $ x>0$ and $\mathrm{sgn}(x)=-1$ if $x<0$.

Finally, we sample a batch of samples $\mathcal{B}$ from the dataset and update YOLOR using gradient back-propagation w.r.t. the loss:
\begin{equation}
\mathcal{L}  =  \frac{1}{\mathcal{B}} \sum_{\mathcal{B}} \left( \mathcal{L}_{ce} + \alpha \cdot \mathcal{L}_{gbpr} \right),
\end{equation}
where $\alpha$ is the coefficient to balance the two losses.

\section{Experiments}
To validate the superior performance of YOLOR, we conducted extensive offline experiments on the Meituan dataset and verified the superiority of YOLOR in online A/B tests. In this section, we first introduce the experimental setup, including the dataset and baseline. Then, in Section \ref{exp_result}, we present the results and analysis of various reranking methods in both offline and online A/B tests.

\subsection{Experimental Setup}
\subsubsection{Dataset}
In order to verify the effectiveness of YOLOR, we conduct sufficient experiments on both public dataset and industrial dataset. For public dataset, we choose Taobao Ad dataset. For industrial dataset, we use real-world data collected from Meituan food delivery platform. Table \ref{tab:my_table} gives a brief introduction to the datasets.

\begin{table}[H]
\caption{Statistics of datasets.}
\label{tab:my_table}
\begin{tabular}{cccc}
\hline
\textbf{Dataset} & \textbf{\#Users} & \textbf{\#Items} & \textbf{\#Records} \\ \hline
Taobao Ad               & 1,141,729        & 99,815           & 26,557,961           \\
Meituan          & 5,648,310      & 14,054,691       & 161,247,488        \\ \hline
\end{tabular}
\end{table}

\begin{itemize}[leftmargin=*]
\item  Taobao Ad \footnote{https://tianchi.aliyun.com/dataset/56}. It is a public dataset collected from the display advertising system of Taobao. This dataset contains more than 26 million interaction records of 1.14 million users within 8 days. Each sample comprises five features: user ID, timestamp, behavior type, item brand ID, and category ID. It includes four behavior types: browse, cart, like, and buy, and each behavior is timestamped. We use the first 7 days as training samples (20170506-20170512), and the 8th day as test samples (20170513).
\item Meituan. It is an industrial dataset collected from the Meituan food delivery platform during August 2024, which contains 161 million interaction records of 5.6 million users within 15 days. The dataset includes 239 features, two labels: click and conversion, and collects all items on the same page as one record. We use the data of the first 14 days as the training set, and the data of the last 1 day as the test set.
\end{itemize}

Note that all samples are list-level, that is, each sample contains all items in an exposed list. We filter out samples whose labels are all 0 or all 1.

\subsubsection{Baseline}
The following six state-of-the-art reranking methods are chosen for comparative experiments and divided into three groups. We select DNN and DeepFM as point-wise baselines (Group I), PRM and MIR as one-stage generator-based baselines (Group II), and Edge-Rerank and PIER as two-stage evaluator-based baseline methods (Group III). A brief introduction of these methods is as follows:

\begin{itemize}[leftmargin=*]
\item $\textbf{DNN}$\cite{dnn} is a basic deep learning method for CTR prediction, which applies MLP for high-order feature interaction.
\item $\textbf{DeepFM}$\cite{deepfm} is a general deep model for recommendation, which combines a factorization machine component and a deep neural network component.
\item $\textbf{PRM}$\cite{prm} adjusts an initial list by applying the self-attention mechanism to capture the mutual influence between items.
\item $\textbf{MIR}$\cite{mir} learns permutation-equivariant representations for the inputted items via self-attention.
mechanism to capture the mutual influence between items.
\item $\textbf{Edge-Rerank}$\cite{gong2022real} generates the context-aware
sequence with adaptive beam search on estimate scores.
\item $\textbf{PIER}$\cite{pier} applies hashing algorithm to select top-k
candidates from the full permutation based on user interests.
\end{itemize}

\subsubsection{Evaluation Metrics. }
We adopt several metrics, i.e., \textbf{AUC} (Area Under ROC Curve) and \textbf{HR} (Hit Ratio) to evaluate YOLOR in offline experiments. To make AUC more suitable for reranking models, we adapt AUC as \textbf{GAUC}\cite{din} (average intra-list AUC) in our experiments. 
AUC measures the global estimated accuracy, and GAUC measures the estimated accuracy within the list. High values in both AUC and GAUC indicate that the model excels at ranking positive samples in front of negative samples, demonstrating strong discriminative power across different evaluation contexts. In online experiments, we adopt CTR and GMV as evaluation metrics.

We use \textbf{HR (Hit Ratio)} \cite{alsini2020hit} to evaluate the consistency of the two stages. For each data, HR is 1 only when the permutations selected by GSU contain the best permutation. Obviously, the HR metrics are only meaningful with evaluator-based reranking methods. The results of HR can be seen in Section \ref{section:consistency}.

It is worth noting that AUC/GAUC and HR can measure two aspects of two-stage methods. AUC/GAUC measures the model's ability to evaluate an ordered list, while HR measures the consistency of the two stages. The deficiency of any indicator will reduce the recommendation effect.
For example, when the reranking model directly returns the result of the ranking stage, the HR will be up to 1 but the AUC will decrease. When the reranking model is very complex, the AUC will increase but the increased time-cost will lead to a decrease in the HR.

\subsubsection{Implementation Details}
We implement all the deep learning baselines and YOLOR with TensorFlow 1.15.0 using NVIDIA A100-80GB GPU. For all comparison models and our YOLOR model, we adopt Adam as the optimizer with the learning rate fixed to 0.001 and initialize the model parameters with normal distribution by setting the mean and standard deviation to 0 and 0.01, respectively. The batch size is 1024, the embedding size is 8. The hidden layer sizes of MLP in Eq. \ref{eq:4} are (1024, 256, 128). 
For the Taobao Ad dataset, the length of the ranking list and reranking list are both 5, thus the length of full permutation is 120. For Metuan dataset page, we select 8 items from the initial ranking list which contains 8 items, thus the length of full permutation is $A_{8}^8=40,320$. For the baseline methods, we follow the settings in PIER and set the number of candidate lists to 100. Similarly, we follow the edge settings and set the beam size to 3.
All experiments are repeated 5 times and the averaged results are reported.

\subsection{Overall Performance}\label{exp_result}

Here we show the results of our proposed method YOLOR. All results are averaged from 5 experiments. As can be seen in Table \ref{tab:2} and Table \ref{tab:3}, YOLOR outperforms baselines including recent two-stage evaluator-based reranking methods.
We have the following observations from the experimental results: 
i) All re-ranking listwise model (e.g. PRM, MIR) makes great improvements over point-wise model (e.g. DNN, DeepFM) by modeling the mutual influence among contextual items, which verifies the impact of context on user clicks behavior.
ii) Compared with generator-based methods(e.g. PRM, MIR), evaluator-based methods also improve the CTR prediction because they evaluate more candidate lists. 
iii) Our proposed YOLOR brings 0.0035/0.0047 absolute AUC and 0.0113/0.0111 absolute GAUC on Taobao/Meituan dataset gains over the state-of-the-art independent baseline which is a significant improvement in industrial recommendation system.

\begin{table}[h]
\caption{Comparison between YOLOR and baseline methods on the Taobao Ad dataset. The best and second-best results in each column are in bold and underlined.}
\label{tab:2}
\begin{tabular}{l|ccc}
\toprule
Model & AUC & GAUC & Loss   \\
\midrule
DNN         & 0.5869 & 0.8130 & 0.1878   \\
DeepFM      & 0.5891 & 0.8132 & 0.1866   \\
PRM         & 0.6152 & 0.8163 & 0.1842   \\
MIR         & 0.6147 & 0.8169 & 0.1853   \\
Edge-Rerank & 0.6286 & 0.8201 & 0.1781\\
PIER        & \underline{0.6316} & \underline{0.8210} & \underline{0.1758}\\
YOLOR       & \textbf{0.6351} & \textbf{0.8323} & \textbf{0.1743}\\
\bottomrule
\end{tabular}
\end{table}

\begin{table}[h]
\caption{Comparison between YOLOR and baseline methods on Meituan. The best and second-best results in each column are in bold and underlined.}
\label{tab:3}
\begin{tabular}{l|ccc}
\toprule
Model & AUC & GAUC & Loss   \\
\midrule
DNN         & 0.7347 & 0.7418 & 0.1162 \\
DeepFM      & 0.7392 & 0.7442 & 0.1154 \\
PRM         & 0.7573 & 0.7595 & 0.1108 \\
MIR         & 0.7598 & 0.7603 & 0.1101 \\
Edge-Rerank & 0.7586 & 0.7605 & 0.1097\\
PIER        & \underline{0.7622} & \underline{0.7638} & \underline{0.1068}\\
YOLOR       & \textbf{0.7669} & \textbf{0.7749} & \textbf{0.1032}\\
\bottomrule
\end{tabular}
\end{table}

\subsection{Consistency Analysis}\label{section:consistency}
\begin{figure}[htbp]	
	\subfigure[HR on Taobao Ad] 
	{
		\begin{minipage}{0.45\linewidth}
			\includegraphics[scale=0.35]{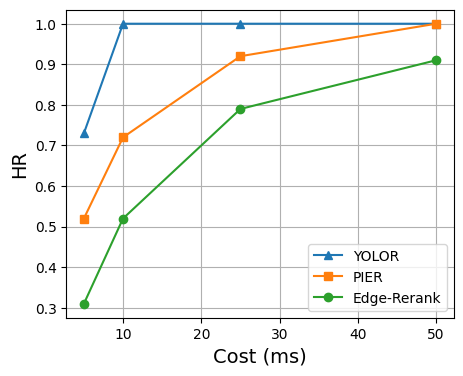}   
		\end{minipage}
	}
	\subfigure[HR on Meituan] 
	{
		\begin{minipage}{0.45\linewidth}
			\includegraphics[scale=0.35]{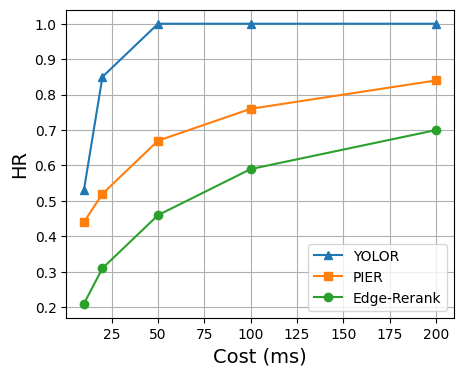}   
		\end{minipage}
	}
\caption{HR results of YOLOR and baseline methods under different cost conditions on Taobao Ad and Meituan datasets.} 
\label{fig:4}  
\end{figure}

We compared the HR (Hit Ratio) of Edge-Rerank, PIER and YOLOR under different inference times on Taobao Ad and Meituan datasets (only evaluator-based methods have HR metrics). All experiments were performed on NVIDIA A100-80GB GPU, with the batch size set to 1024 and averaged 100 times. In the Meituan data set, there are a total of $A_{8}^8=40,320$ candidate lists. We carefully adjust the number of candidate lists of the model to make the model time-consuming equal, and keep the error within 1 ms. For YOLOR, we use random generation to conduct experiments. 
As shown in Figure \ref{fig:4}, we have the following observations: i) YOLOR performs well across all periods. This is because YOLOR is a very efficient method that is capable of retrieving more candidate lists at the same time. ii) As time increases, the HR of all methods improves, since the amount of retrieval lists for all methods becomes larger. iii) YOLOR can traverse all candidate lists within 50ms, achieving an HR of 1.

\subsection{Ablation Study}
To assess the effectiveness of each component in YOLOR, we conducted a series of ablation studies using the Taobao Ad and Meituan datasets. Specifically, we build several variants of the YOLOR:
\begin{itemize}[leftmargin=*]
\item w/o IRM. A variant of YOLOR without the IRM, which means only using candidate items' raw embedding  $\mathbf{X}$  instead of the semantic embedding $\mathbf{X}^s$.
\item w/o TCEM. A variant of YOLOR without the TCEM, which is replaced by a single global self-attention.
\item w/o GBPR. A variant of YOLOR without the GBPR loss in Eq. \ref{eq:gbpr}.
\end{itemize}

\begin{table}[h]
\caption{The contributions of different components of YOLOR.}
\label{tab:5}
\begin{tabular}{l|cc|cc}
\toprule
\multirow{2}{*}{Model} & \multicolumn{2}{c|}{Taobao Ad} & \multicolumn{2}{c}{Meituan}  \\
 & AUC & GAUC   & AUC & GAUC \\
\midrule
w/o IRM           & 0.5712 & 0.7940 & 0.7332 & 0.7454  \\
w/o TCEM           & 0.6236 & 0.8225 & 0.7574 & 0.7621  \\
w/o GBPR          & 0.6305 & 0.8207 & 0.7629 & 0.7632  \\
YOLOR             & \textbf{0.6351} & \textbf{0.8323} & \textbf{0.7669} & \textbf{0.7749} \\
\bottomrule
\end{tabular}
\end{table}

Table \ref{tab:5} shows the results of the ablation study, and we can draw the following conclusions: 1) without the IRM, the model's performance decreases significantly because accurate point-wise predictions are the basis of YOLOR; 2) without TCEM, the performance of the model declines, indicating that TCEM has stronger context extraction capabilities comparing with single self-attention; 3) without GBPR, the performance of the model declines, indicating that the intra-list loss is valid for reranking.

To assess the efficiency of CCM, we remove the CCM module and randomly sample K candidate lists to evaluate their HR and time-consuming. And we set different K values for detailed comparison.

\begin{table}[h]
\caption{HR and Cost of different sample numbers without CCM.}
\label{tab:52}
\begin{tabular}{l|cc|l|cc}
\toprule
\multirow{2}{*}{Settings
} & \multicolumn{2}{c|}{Taobao Ad} &\multirow{2}{*}{Settings
}  & \multicolumn{2}{c}{Meituan}  \\
 & HR & Cost (ms) &  & HR & Cost (ms)\\
\midrule
K=5            & 0.0417 &  4.7 & K=100 & 0.0025 &  56.1  \\
K=10           & 0.0833 &  7.1 & K=200 & 0.0050 & 103.4  \\
K=20           & 0.1667 & 11.8 & K=300 & 0.0074 & 169.8  \\
K=50           & 0.4167 & 28.6 & K=400 & 0.0099 & 217.7  \\
\bottomrule
\end{tabular}
\end{table}
As shown in Table \ref{tab:52}, after the CCM module is removed, all vectors in both IRM and TCEM must be recalculated, which significantly increases the average time cost. Furthermore, due to random sampling constraints, the Hit Ratio (HR) shows a marked decline as it is sensitive to the number of evaluated candidate lists.


\subsection{Hyperparameter Analysis}
We analyze the impact of weight $\alpha$ in GBPR loss. Table \ref{tab:Hyperparameter} shows the results of our experiments and we can find that $\alpha$ significantly affects YOLOR’s AUC/GAUC metric. As $\alpha$ increases, AUC first increases and then decreases while GAUC increases and stabilizes at a high level.

\begin{table}[h]
\caption{The effect of parameter weight $\alpha$ in GBPR loss.}
\label{tab:Hyperparameter}
\begin{tabular}{l|cc|cc}
\toprule
\multirow{2}{*}{Settings} & \multicolumn{2}{c|}{Taobao Ad} & \multicolumn{2}{c}{Meituan}  \\
 & AUC & GAUC   & AUC & GAUC \\
\midrule
$\alpha$=0            & 0.6305 & 0.8207 & 0.7629 & 0.7632  \\
$\alpha$=0.01         & 0.6334 & 0.8297 & 0.7641 & 0.7704  \\
$\alpha$=0.05         & \textbf{0.6351} & \textbf{0.8323} & \textbf{0.7669} & 0.7749 \\
$\alpha$=0.1          & 0.6345 & 0.8323 & 0.7664 & 0.7749  \\
$\alpha$=0.5          & 0.6341 & 0.8323 & 0.7658 & \textbf{0.7750}  \\
\bottomrule
\end{tabular}
\end{table}

\subsection{Performance on Online System} \label{subsection:5.6}
\begin{figure}[h]
\centering
\includegraphics[width=0.9\linewidth, height=1\textheight, keepaspectratio]{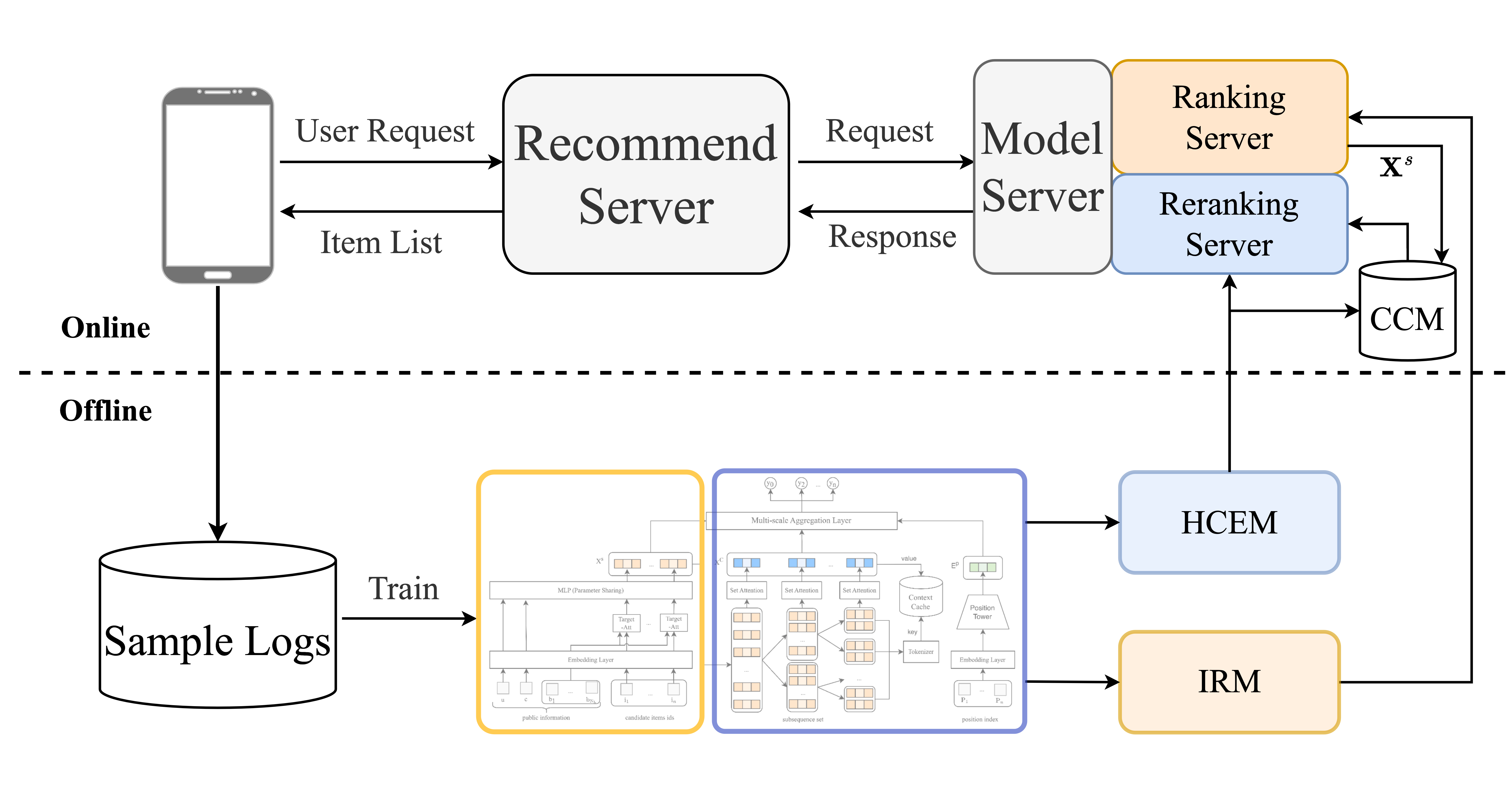}
\caption{Architecture of the online deployment with YOLOR.}
\label{fig:yolor_online}
\end{figure}

To evaluate the online performance of YOLOR, we deployed YOLOR on the Meituan Shichiguan business as shown in Figure \ref{fig:yolor_online}.


   

We also conducted a rigorous A/B test for three weeks, from March 2025 to April 2025. Specifically, we assigned YOLOR with 30\% traffic, while the remaining 70\% traffic was assigned to baseline (PIER).
Table \ref{tab:online} shows the online performance of YOLOR. Compared to the baseline model (PIER), YOLOR has increased the CTR by 5.13\% and the GMV by 7.64\%, which are very significant growth to business.
Besides, we find that time consumption has not increased at all, which is an important indicator to determine whether it can be applied to large-scale industrial scenarios.
Now, YOLOR is deployed on the Meituan food delivery platform and serves millions of users.

\begin{table}[h]
\caption{Online A/B test result.}
\label{tab:online}
\begin{tabular}{ccccc}
\hline
\textbf{Method}        & \textbf{CTR} & \textbf{GMV}  & \textbf{Cost (ms)} & \textbf{Time-out} \\ \hline
\textbf{YOLOR}        &  +5.13\%  & +7.64\%  & -0.003   & -0.001\%  \\
\hline
\end{tabular}
\end{table}

\section{Conclusion}
In this paper, we identify the inherent trade-off issues in two-stage re-ranking methods, which cannot simultaneously balance efficiency and effectiveness. To overcome this limitation, we propose a tree-based approach, named YOLOR, designed to achieve both "permutation-level efficiency" and "list-level effectiveness."
Specifically, we design a Tree-based Context Extraction Module (TCEM) that integrates multi-scale contextual information to accurately evaluate the list. Additionally, we develop a Context Cache Module (CCM) to enable efficient reuse of multi-scale contextual information across lists. The efficiency of YOLOR allows you to only evaluate once.
Both offline experiments and online A/B tests show that YOLOR significantly outperformed other existing reranking baselines. We have deployed YOLOR on the Meituan food delivery platform.

\bibliographystyle{ACM-Reference-Format}
\bibliography{main}

\end{document}